\documentclass[manuscript,screen,nonacm,10pt]{acmart}
\usepackage{caption}
\usepackage{subcaption}
\usepackage{amsmath}
\AtBeginDocument{%
  \providecommand\BibTeX{{%
    \normalfont B\kern-0.5em{\scshape i\kern-0.25em b}\kern-0.8em\TeX}}}

\begin{document}

\title[Streaming Chain]{Streaming Chain}

\author{Yi Lyu}
\authornotemark[1]
\affiliation{%
  \institution{University of Wisconsin-Madison}
  \city{Madison}
  \state{WI}
  \country{USA}}
\email{ylyu76@wisc.edu}


\newcommand{\BC}{\texttt{BlockCreation} }

\begin{abstract}
\normalsize
Blockchain and blockchain-inspired decentralized applications are on the rise thanks to their unique characteristics such as their decentralized nature, anonymity, tamper-proof, etc. However, blockchain transactions tend to experience long end-to-end latency. A major contributor to this latency is the block creation step, which might block transaction processing. There are two approaches to ameliorate the block creation overhead: (1) speed up the block creation process and (2) process transactions before block creation finishes. In this project, we work towards designing a self-designed adaptive block creation process that automatically selects the optimal configurations based on the workload and hardware resources by (1) defining mathematical models to predict transaction latency based on design and environmental parameters; (2) developing measurement techniques to collect performance-related metrics in docker-hosted blockchain systems and observing trends to build intuition; (3) defining mathematical model to predict transaction success rate under various key accessing patterns and block size configurations, and validating the model with simulation-based measurements.
\end{abstract}





\maketitle
\section{Overview}
On a private (i.e., permissioned) blockchain\cite{zhang2021tapping, zhang2023first, feng2021allign, feng2025optimus, feng2024f3, cheng2024vetrass, han2022francis}, the lifecycle of a transaction typically consists of the following steps: (1) a client application submits a signed transaction proposal, (2) peer(s) of organization(s) execute the transaction and sign the transaction proposal response, (3) an orderer receives the signed transaction, orders it, verifies its signature, and adds it into a block of ordered transactions, and (4) peers receive an ordered transaction block, verify the signatures, status, and endorsements of transactions, and then commits the transactions to the ledger.
\subsection{Delay on the Orderer Side}
Although the workflow sounds straightforward, step (3) where the orderer creates a block of ordered transactions turns out to be a bottleneck of fast transaction commitment. Orderers wait until they receive enough transactions or the maximum elapsed duration for a block is met before they start the ordering process. Such a design introduces a transaction queuing delay on the orderer side. To reduce the waiting time, a simple solution would be to decrease the desired number of transactions per block. However, this would require both the orderer and the peers to pay the cost of hashing block content more frequently. Depending on the available computational resources, the hash computation can be expensive and in turn, slows down the end-to-end transaction processing. Having smaller blocks also means distributing more blocks from the orderer to peers. This slows down the communication phase by increasing the volume of transmitted data (due to the header fields overhead), network stack traversals, and packet re-transmissions when the network connection is unstable.

With the aforementioned tradeoffs in mind, we aim to design and implement a self-designed adaptive block creation process that achieves fast transaction processing regardless of application scenarios.

\subsection{Current progress}
Our current effort is summarized by:
\begin{enumerate}
\item defining mathematical models to predict transaction latency based on design and environmental parameters;
\item developing measurement techniques to collect performance-related metrics in docker-hosted blockchain systems and observing trends to build intuition\cite{catp};
\item defining mathematical model to predict transaction success rate under various key accessing patterns and block size configurations, and validating the model with simulation-based measurements.
\end{enumerate}

\section{Block Creation Modeling}
\subsection{Design}
Table \ref{tab:ep_dp} shows environmental parameters and design parameters in blockchain calculator.  

\begin{table}[H]
  \caption{Summarization of Environmental Parameters (EP) and Design Parameters (DP) in the block creation
  design continuum.}
  \tiny
  \label{tab:ep_dp}
\begin{tabular}{l|l|l|l|l|l}
 & Variable & Name & Description & Value & Unit \\ \hline
EP & $R$ & Transaction arrival rate & Average time between transaction arrivals to the orderer & $(0, \infty)$ & seconds \\
 & $TS$ & Transaction size & Size of one transaction, depending on the schema & $[1, \infty)$ & bits \\
 & $NB$ & Network bandwidth & Data transfer speed through order-peer network links & $[1, \infty)$ & bit/second \\
 & $DB$ & Disk I/O bandwidth & Disk I/O bandwidth & $[1, \infty)$ & bit/second \\
 & $CS$ & CPU speed & Characterize the orderer CPU execution speed & $[1, \infty)$ & cycles/second \\ \hline
DP & $BS$ & Batch size & Max number of transactions in a block & $[1, \infty)$ & \# transactions \\
 & $BTO$ & Batch time out & Max time to wait for next transaction before block creation & $(0, \infty)$ & seconds \\
 & $SF$ & Signature function & Hash function used to compute block signature & SHA256, MD5 & categorical
\end{tabular}
\end{table}

\subsubsection{I/O Time Modeling}

The I/O time for one block creation can be modeled as following, where $m_0$ is the metadata size in bits for each block\cite{vetrass}.
\begin{equation}
    \sigma_{I/O} = \frac{m_0 + BS \times TS}{DB} + \frac{m_0 + BS \times TS}{NB}
\end{equation}

\subsubsection{CPU Time Modeling}
The CPU time for one block creation is mostly spent on calculating hash for the block data using signature function \texttt{SF}.  We choose to model the CPU cycles cost by signature function to be linear to the block data size, which is true for \texttt{SHA256} used in the \texttt{hyperledger-fabric} project.

The CPU time can be modeled as following, where $c_0$ is the CPU cycles spent in other parts (e.g. calculating hash on block header) and $c_1$ is the average CPU cycles spent in signature function for each bit of the block data. 

\begin{equation}
    \sigma_{cpu} = \frac{c_0 + c_1 \times BS \times TS}{CS} 
\end{equation}

\subsubsection{Transaction Latency Modeling}
The average transaction latency can be modeled as the equation below. $\frac{\min(BTO, \frac{BS}{R})}{2}$ is the average time for one transaction to wait before block generation.
\begin{equation}
    \sigma_{latency} = \frac{\min(BTO, \frac{BS}{R})}{2} + \sigma_{I/O} + \sigma_{cpu}
\end{equation}
\subsection{Evaluation}
{\bf Experiment setup:}  

{\em Hardware. } We use a server with 20-core Intel Xeon Sliver 4114 CPU, 128GB DDR4 RAM and 1TB HDD.  

{\em Topology. } We use the \texttt{test-network} \cite{testNetwork} in Hyperledger Fabric to test a topology of two peer nodes and one orderer node\cite{monom}. Each peer node belongs to a different peer organization and the single orderer node uses Raft to provide ordering service. All nodes in \texttt{test-network} are deployed as docker containers on a single server. 

{\em Workload. } We use YCSB workload with 1000 update operations (transactions). And we use default configurations for other parameters in YCSB workload (e.g. ten 100-byte fields for each record). Our benchmark code is based on previous blockbench paper \cite{dinh2017blockbench, Efflex} and we open source it on Github \cite{blockbench}. The transaction arrival distribution is uniform distribution and we configure the transaction arrival rate to be $\{8, 16, 32\}$ transactions per second (txn/s).

{\em Model. } We use the transaction latency model discussed in previous section. And based on our previous analysis, $\sigma_{I/O}+\sigma_{cpu}$ can be modeled using a linear model $c_0 \cdot BS + c_1$ ($BS$ is maximum number of transactions in one block). And we use linear regression provided by Python package \texttt{scikit-learn} to determine $c_0$ and $c_1$.
\begin{equation}
\begin{split}
\sigma_{latency}  & = \frac{\min(BTO, \frac{BS}{R})}{2} + \sigma_{I/O} + \sigma_{cpu}\\
& = \frac{\min(BTO, \frac{BS}{R})}{2} + c_0 \cdot BS + c_1 
\end{split}
\end{equation}

{\bf Experiment results:}  

\begin{figure}[!htp]
    \centering
    \includegraphics[scale=0.4]{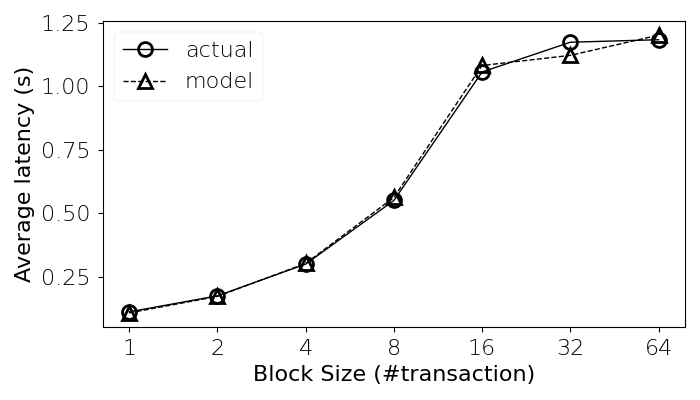}
    \caption{Accurately predicting average transaction latency (8 transactions/s)}
    \label{fig:blk_latency_exp1}
\end{figure}

{\em Low transaction arrival rate.} We vary the block size $BS$ and measure the average transaction latency under transaction arrival rate 8 txn/s. Figure \ref{fig:blk_latency_exp1} shows the average transaction latency from actual measurement and model estimation\cite{safeguard}.  First, our transaction latency model captures the trend that average transaction latency increases with larger block size. Second, our transaction latency model is accurate that over-estimation error is less than $2.4\%$ and under-estimation error is less than $4.5\%$. 

\begin{figure}[!htp]
    \centering
    \includegraphics[scale=0.4]{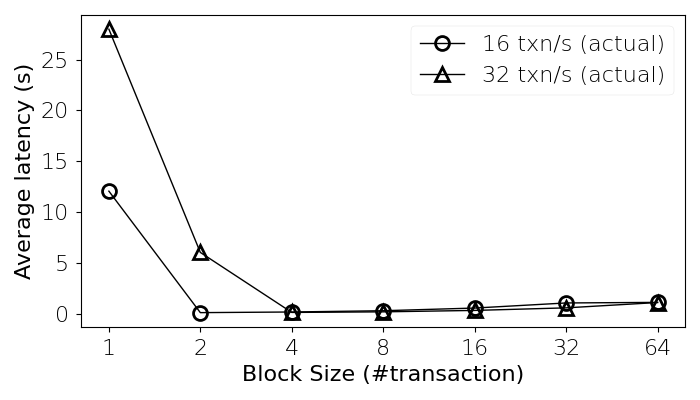}
    \caption{Average transaction latency under high transaction arrival rate}
    \label{fig:blk_latency_exp2}
\end{figure}

{\em High transaction arrival rate.} We vary the block size $BS$ and measure the average transaction latency under transaction arrival rate $\{16, 32\}$ txn/s. Figure \ref{fig:blk_latency_exp2} shows the average transaction latency from actual measurement.  

The average transaction latency trend is 
different from {\em low transaction arrival rate} case. With small block size, the average transaction latency is much larger than other block size settings. For example, the average transaction latency is 11.78s under $BS=1$ and 16 txn/s, which is more than 19X higher than other $BS$ settings. \textbf{Our current transaction latency model can not capture this trend}. 

We think the above behavior is related to the max block processing rate $BP\_RATE$ in the \texttt{test-network} system. And if transaction arrival rate $R$ is larger than the product of $BS$ and $BP\_RATE$ ($R > BS \cdot BP\_RATE$), there is extra queuing delay before transactions can be put into the block. 

Our initial investigation on \texttt{hyperledger} systems shows that the block processing time for smallest block ($BS = 1$) is on average 84.32 ms in peer side, which translates to $BP\_RATE=11.85$ blocks per second (blk/s). When $BS=1$ and $R=16$, $R > BS \cdot BP\_RATE$ holds and there exists queuing delay for transactions.
And the detailed block generation time breakdown is shown in Table \ref{tab:bs_breakdown}. State validation represents the process of validating transactions in blocks. Block and private data commit represents the process of  persisting the newly generated block on disk. State commit represents the process of executing transactions in blocks.  History DB presents the process of updating history database.  
\begin{table}[!htp]
    \centering
    \begin{tabular}{c|c}
        Component & Time (ms) \\
        \hline
        State validation & 0.09  \\
        \hline
        Block and private data commit & 49.83 \\ 
        \hline
        State commit & 15.75 \\ 
        \hline
        History DB & 15.79 \\ 
        \hline
    \end{tabular}
    \caption{Block processing time breakdown on peer side ($BS$=1)}
    \label{tab:bs_breakdown}
\end{table}

We leave the queuing delay part $\sigma_{queue}$ as future work to be added into our current transaction latency model.

\section{Resource Usage Evaluation}
To estimate the resource usage of choosing different block sizes under a given transaction arrival rate, we also collect resource usage number of docker containers during each run of the YCSB workload.
\subsection{Docker Performance Metrics}
 To get the CPU usage of each peer container and the ordered container, we first get the PID of each container through "docker top" ~\cite{DockerTop}, and then periodically read the "/proc/[PID]/stat" file to get the amount of time that this process has been scheduled in kernel and user mode; finally, we divide this amount of time by the total amount of time the machine has spent in various states (available in "/proc/stat"). We also collect the amount of data written to and read from block devices (BlockIO) and the amount of data received and sent over the host network interface (NetIO) of docker containers with "docker stats" ~\cite{DockerStats}. 
 We show graphs of resource usage we gather when running the aforementioned YCSB workload in the following subsections.

\subsection{CPU Percentage}
\begin{figure}[h]

    \centering
    \begin{subfigure}[h]{1.0\textwidth}
    \begin{minipage}[h]{0.48\linewidth}
      \centering 
        \includegraphics[trim={0cm 0cm 0cm 0.2cm},clip,width=\textwidth]{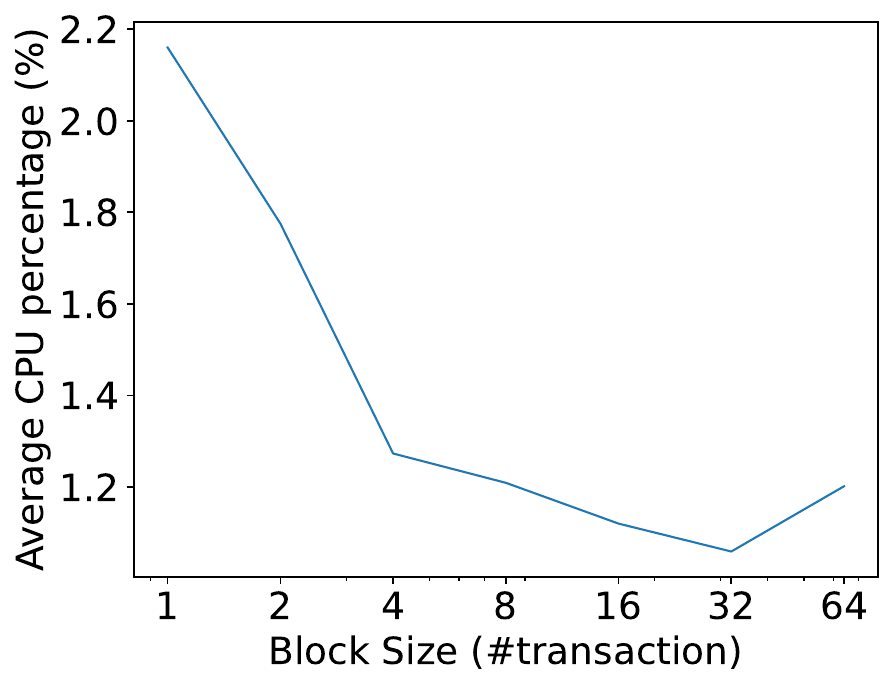}
        \caption{orderer CPU usage}
        \label{fig:cpu_orderer_tx8}
    \end{minipage}
    \begin{minipage}[h]{0.48\linewidth}
      \centering 
    \includegraphics[trim={0cm 0cm 0cm 0.2cm},clip,width=\linewidth]{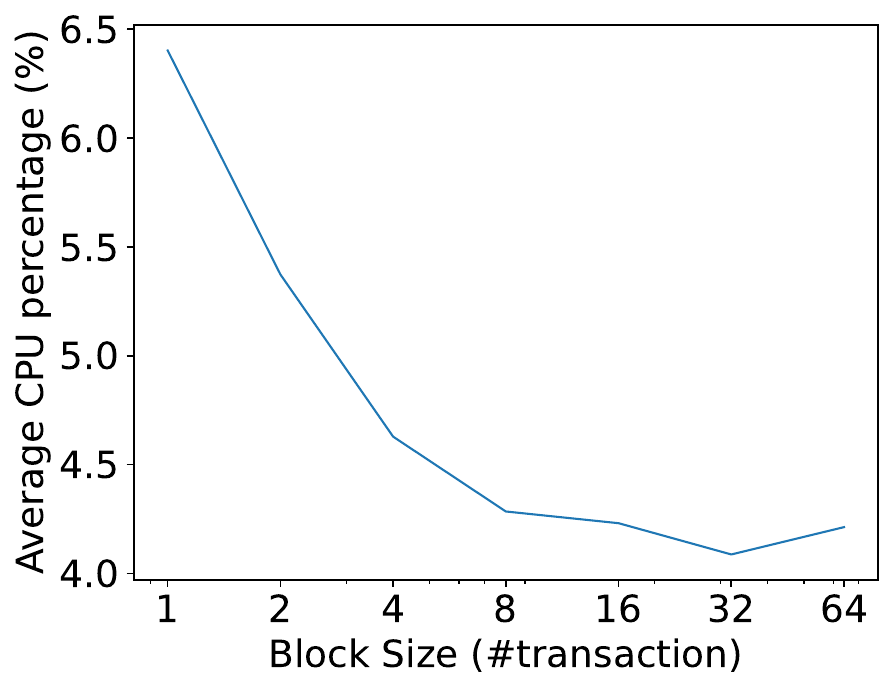}
    \caption{peer CPU usage}
    \label{fig:cpu_peer_tx8}
    \end{minipage}
    
    \end{subfigure}
    \caption{CPU usage percentage graphs when the arrival rate is 8 transactions/s}
    \label{fig:cpu_usage_tx8}  
\end{figure}

{\em Low transaction arrival rate.} As shown in Figure~\ref{fig:cpu_usage_tx8}, when the transaction rate is 8 transactions per second, the CPU usage percentage of the order and peers decreases as we increase the block size (until we reach block size 32). Although the CPU usage is generally low, users can opt for a larger block size if they are very sensitive to their CPU usage cost.

{\em High transaction arrival rate.}
When the transaction rate is high, we observe an overall higher average CPU usage for both the orderer and the peers. Our previous observation that the CPU usage percentage decreases until we get close to 32 transactions per block still holds here.
\begin{figure}[h]

    \centering
    \begin{subfigure}[h]{1.0\textwidth}
    \begin{minipage}[h]{0.48\linewidth}
      \centering 
        \includegraphics[trim={0cm 0cm 0cm 0.2cm},clip,width=\textwidth]{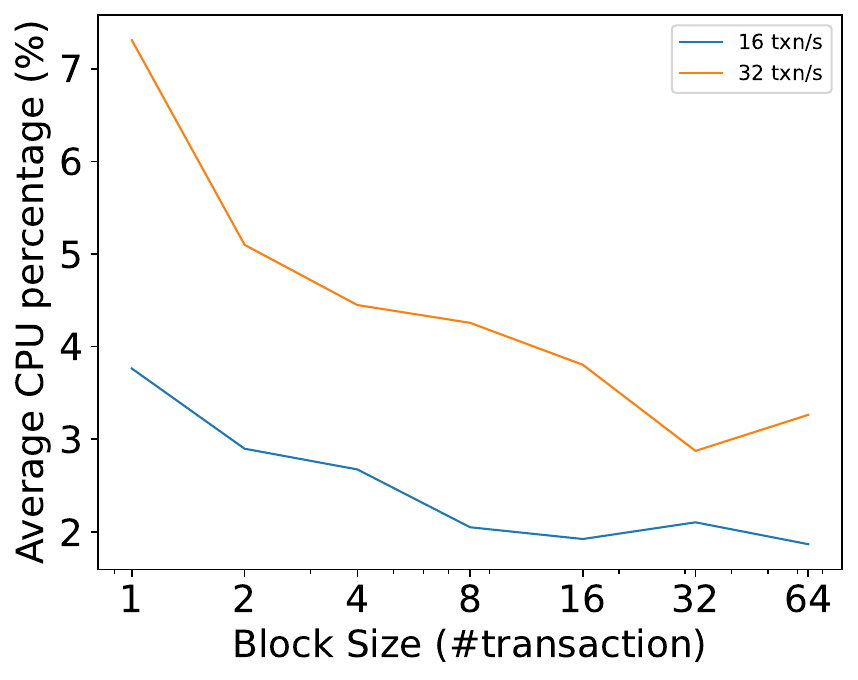}
        \caption{orderer CPU usage}
        \label{fig:cpu_orderer_high_rate}
    \end{minipage}
    \begin{minipage}[h]{0.48\linewidth}
      \centering 
    \includegraphics[trim={0cm 0cm 0cm 0.2cm},clip,width=\linewidth]{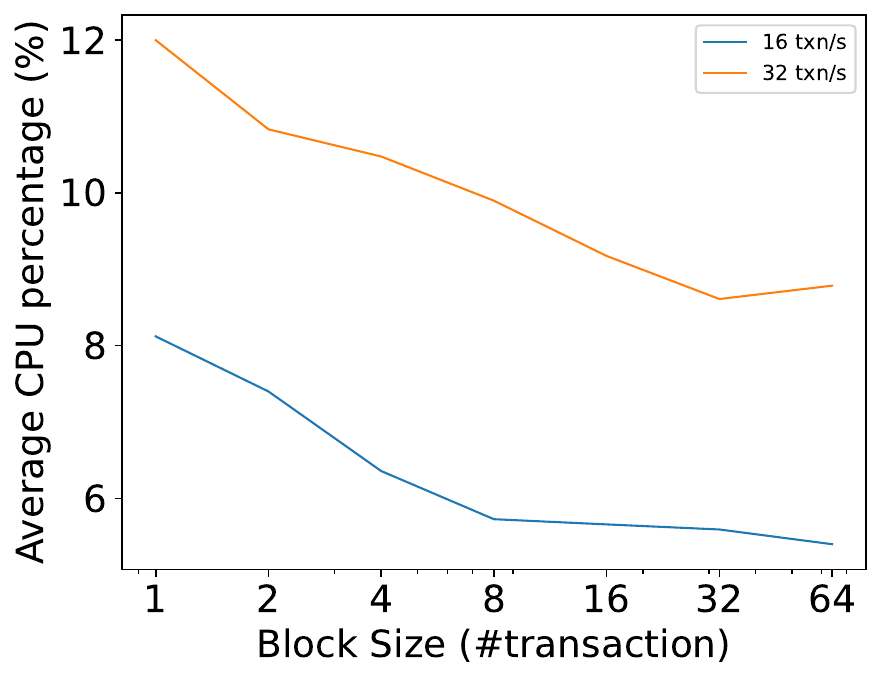}
    \caption{peer CPU usage}
    \label{fig:cpu_peer_high_rate}
    \end{minipage}
    
    \end{subfigure}
    \caption{CPU usage percentage graphs when the arrival rate is 16 or 32 transactions/s}
    \label{fig:cpu_usage_high_rate}  
\end{figure}




\subsection{Block I/O} 
\begin{figure}[h]

    \centering
    \begin{subfigure}[h]{1.0\textwidth}
    \begin{minipage}[h]{0.48\linewidth}
      \centering 
        \includegraphics[trim={0cm 0cm 0cm 0.2cm},clip,width=\textwidth]{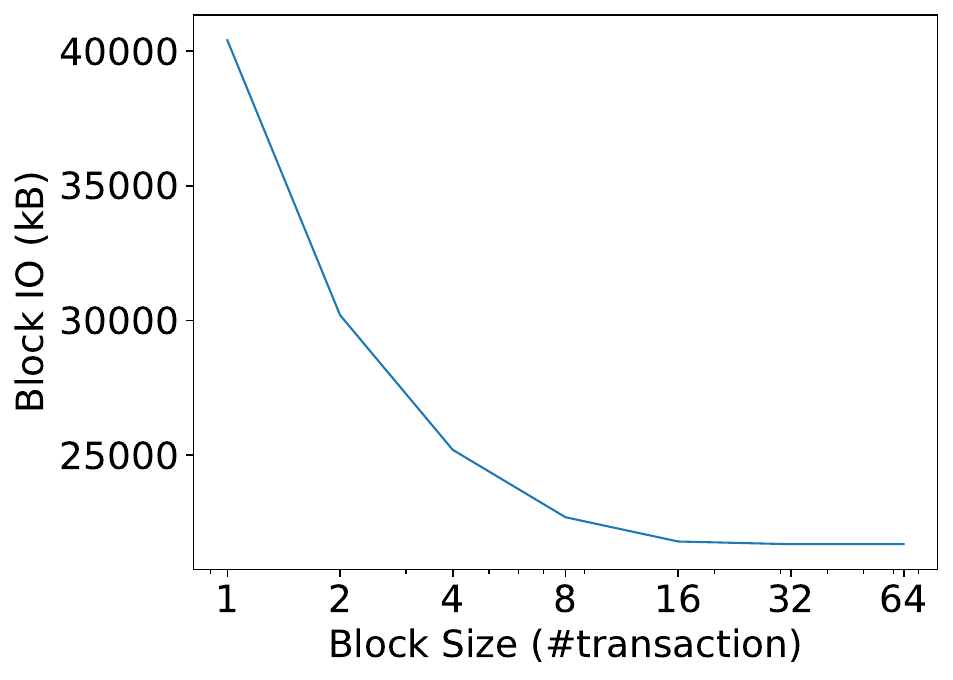}
        \caption{orderer Block IO}
        \label{fig:block_io_orderer_tx8}
    \end{minipage}
    \begin{minipage}[h]{0.48\linewidth}
      \centering 
    \includegraphics[trim={0cm 0cm 0cm 0.2cm},clip,width=\linewidth]{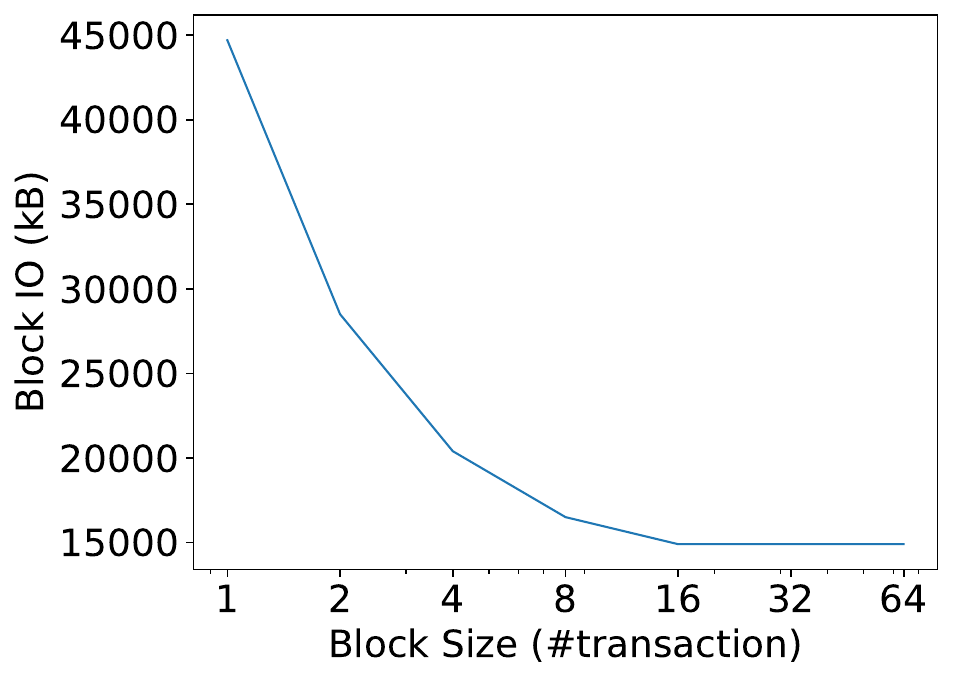}
    \caption{peer Block IO}
    \label{fig:block_io_peer_tx8}
    \end{minipage}
    
    \end{subfigure}

    \caption{Block IO graphs when the arrival rate is 8 transactions/s}

    \label{fig:blockio_tx8}  
\end{figure}

\begin{figure}[h]

    \centering
    \begin{subfigure}[h]{1.0\textwidth}
    \begin{minipage}[h]{0.48\linewidth}
      \centering 
        \includegraphics[trim={0cm 0cm 0cm 0.2cm},clip,width=\textwidth]{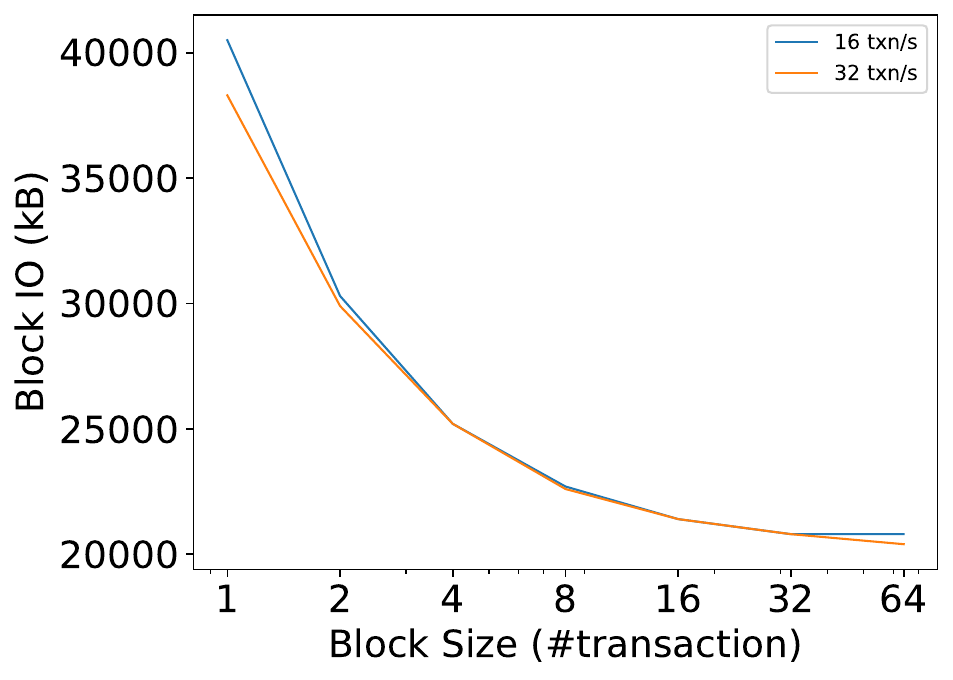}
        \caption{orderer Block IO}
        \label{fig:block_io_orderer_high_rate}
    \end{minipage}
    \begin{minipage}[h]{0.48\linewidth}
      \centering 
    \includegraphics[trim={0cm 0cm 0cm 0.2cm},clip,width=\linewidth]{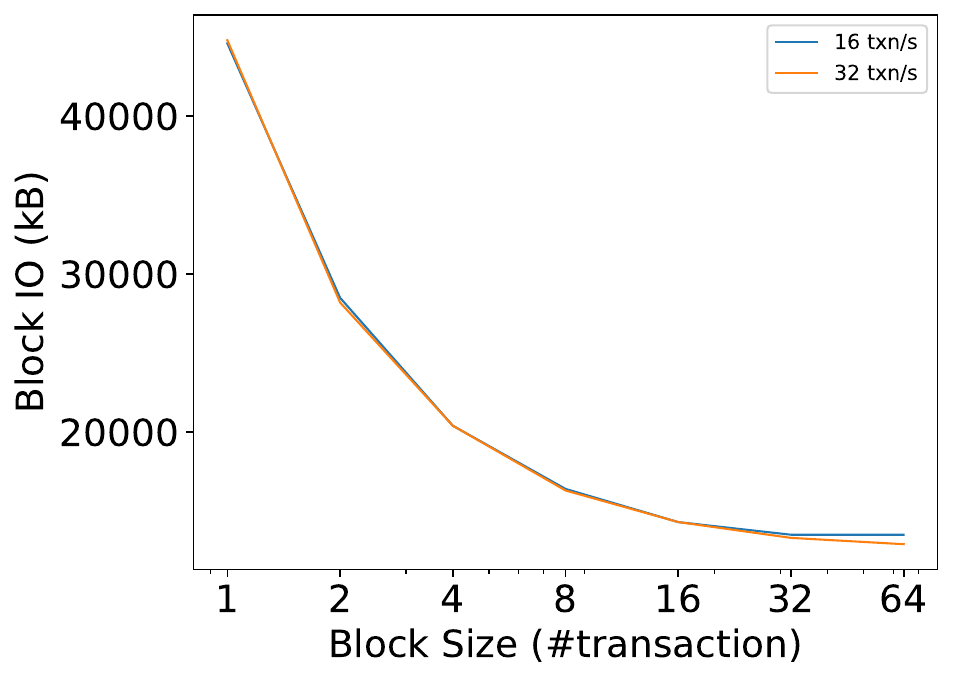}
    \caption{peer Block IO}
    \label{fig:block_io_peer_high_rate}
    \end{minipage}
    
    \end{subfigure}

    \caption{Block IO graphs when the arrival rate is 16 or 32 transactions/s}

    \label{fig:blockio_high_rate}  
\end{figure}

As Figure~\ref{fig:blockio_tx8} and ~\ref{fig:blockio_high_rate} show, the total amount of block IO doesn't change notably as the transaction rate increases. This is expected as different transaction rate shouldn't change the overall volume of transaction data. However, as the number of transactions per block increases, Block IO generally decreases. This is also as expected because having fewer blocks decreases the block metadata cost.

\subsection{Net I/O} 
\begin{figure}[h]

    \centering
    \begin{subfigure}[h]{1.0\textwidth}
    \begin{minipage}[h]{0.48\linewidth}
      \centering 
        \includegraphics[trim={0cm 0cm 0cm 0.2cm},clip,width=\textwidth]{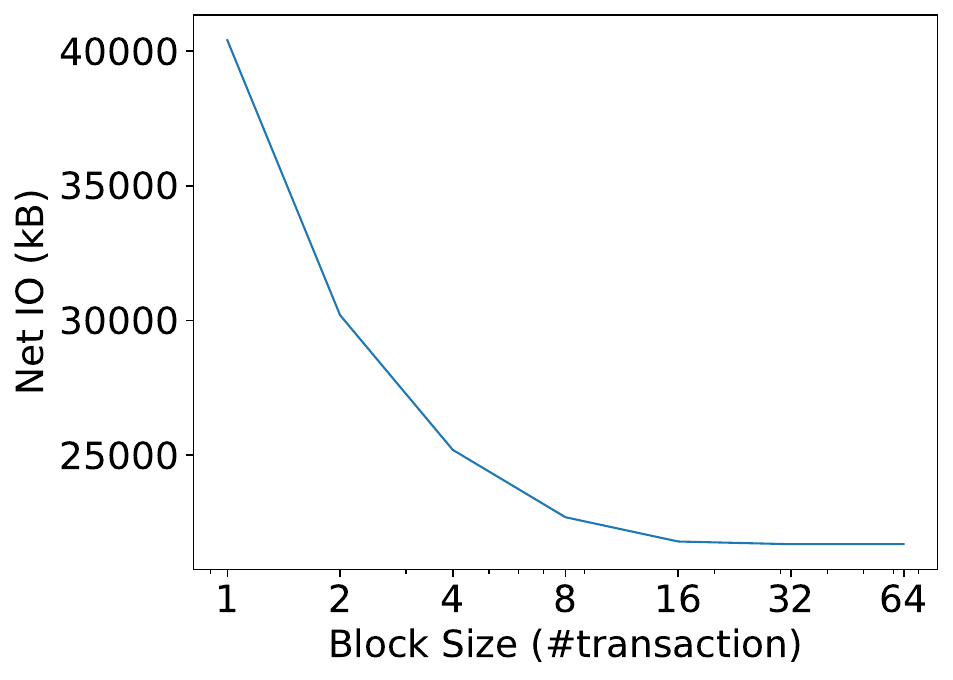}
        \caption{orderer Net IO}
        \label{fig:net_io_orderer_tx8}
    \end{minipage}
    \begin{minipage}[h]{0.48\linewidth}
      \centering 
    \includegraphics[trim={0cm 0cm 0cm 0.2cm},clip,width=\linewidth]{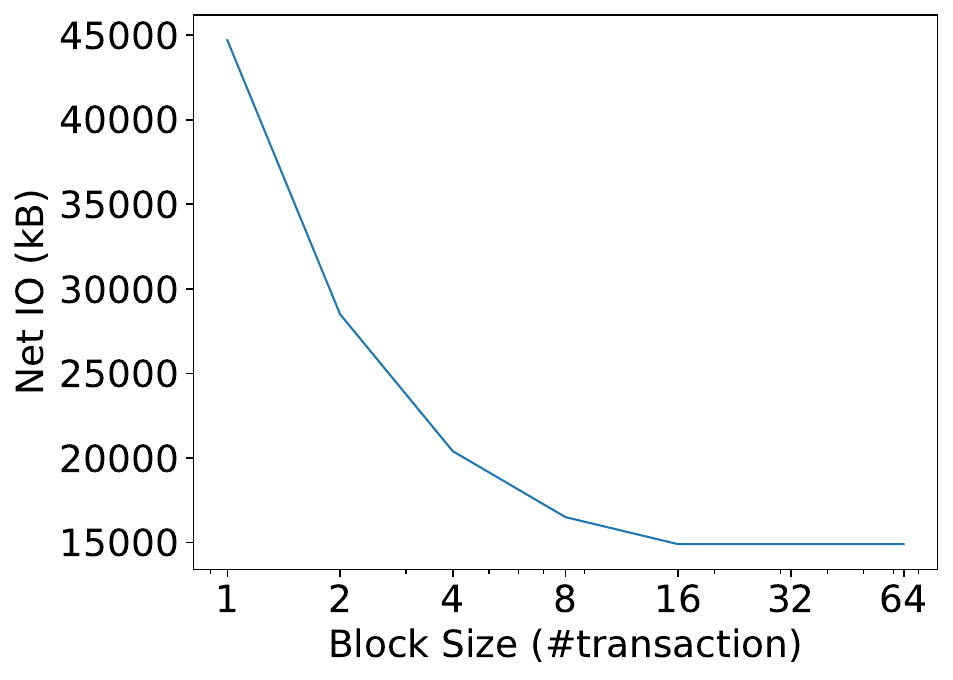}
    \caption{peer Net IO}
    \label{fig:net_io_peer_tx8}
    \end{minipage}
    
    \end{subfigure}

    \caption{Net IO graphs when the arrival rate is 8 transactions/s}

    \label{fig:netio_tx8}  
\end{figure}

\begin{figure}[h]

    \centering
    \begin{subfigure}[h]{1.0\textwidth}
    \begin{minipage}[h]{0.48\linewidth}
      \centering 
        \includegraphics[trim={0cm 0cm 0cm 0.2cm},clip,width=\textwidth]{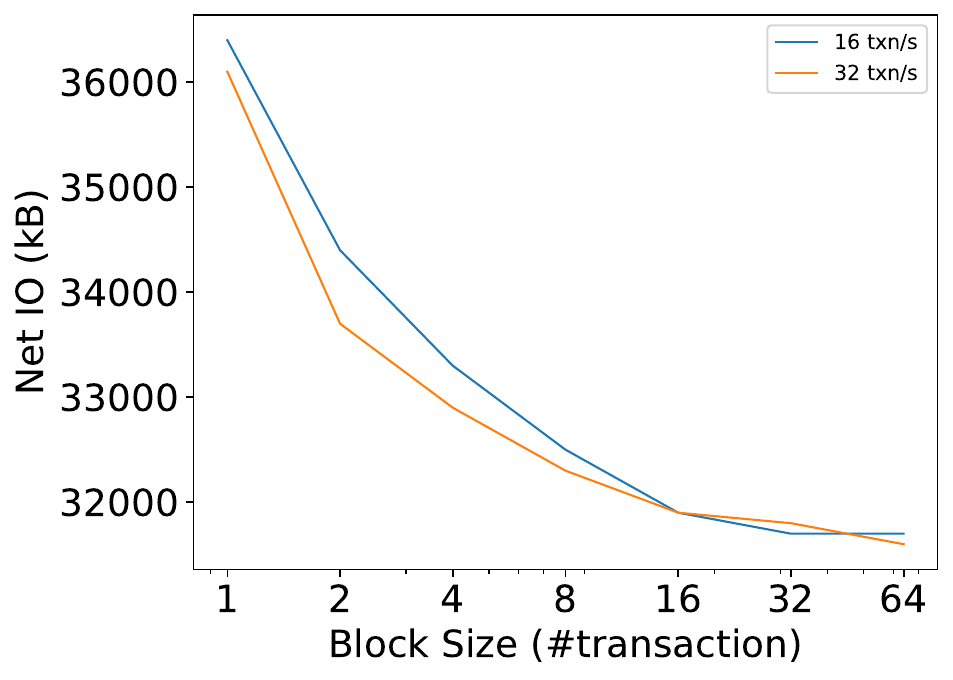}
        \caption{orderer Net IO}
        \label{fig:net_io_orderer_high_rate}
    \end{minipage}
    \begin{minipage}[h]{0.48\linewidth}
      \centering 
    \includegraphics[trim={0cm 0cm 0cm 0.2cm},clip,width=\linewidth]{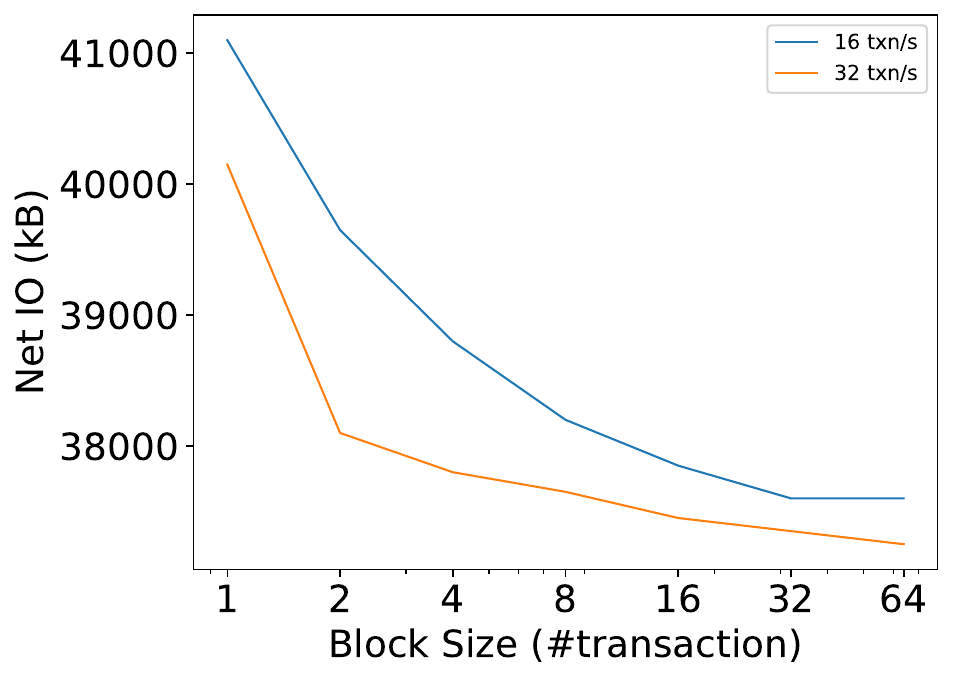}
    \caption{peer Net IO}
    \label{fig:net_io_peer_high_rate}
    \end{minipage}
    
    \end{subfigure}

    \caption{Net IO graphs when the arrival rate is 16 or 32 transactions/s}

    \label{fig:netio_high_rate}  
\end{figure}

Figure~\ref{fig:netio_tx8} and ~\ref{fig:netio_high_rate} first show that for both the orderer and the peers, the total amount of Net IO decreases as the block size increases. As explained above, this is expected because of the reduced metadata and header fields overhead. However, an unexpected observation is that higher transaction rates appear to result in a smaller total amount of Net IO. We suspect that communication between containers on one server is not an accurate reflection of real-world inter-machine communication. Running experiments across multiple machines in the future will give us more meaningful network IO and latency results.

\section{Transaction Success Rate Modeling}

In this section, we focus on theoretical and empirical evaluation of transaction failure probabilities in a blockchain system. First, we define a mathematical model for transaction success rate under different environment (i.e. workload distribution) and design (i.e. block size) parameter configurations. Then, we evaluate its accuracy by comparing the modeled success rate with simulated trials. An accurately defined transaction success rate model will guide a self-designing block creation system to make reliable and robust design choices, especially whether to make the peers wait for block creation to complete.

\subsection{Theoretical model}
\label{sec:theoretical-model}

\textbf{Definition of failure.} In a blockchain-based key-value storage application, each block contains an ordered sequence of transactions. Each transaction represents a user-submitted read/write request to a key. Let $(a, b)$ be two transactions within the same block and $a$ is ordered before $b$, we define the three following failure scenarios:
\begin{itemize}
    \item Read-Write Failure (RWFail): $a$ requests to read a key, and $b$ attempts to write to the same key.
    \item Write-Read Failure (WRFail): $a$ requests to write to a key, and $b$ attempts to read the same key.
    \item Write-Write Failure (WWFail): $a$ requests to write to key, and $b$ attempts to write to the same key.
\end{itemize}

Note that in above failure scenarios, even $b$'s request would fail, $a$'s request could succeed as long as $a$ does not form any failure scenario with any other transaction earlier than $a$.

\noindent\textbf{Assumptions.} Our success rate model assumes (1) uniform access pattern across all clients, including the distribution of read/write requests and the distribution of keys; and (2) consistent, uniform block size in terms of number of transactions within a block; note that in alternative approaches we might model the block size as a probability distribution characterized by transaction arrival rate and block timeout; (3) finite key spaces; and (4) the transactions submitted by different peers/clients are i.i.d. events.

\noindent\textbf{Inputs.} Table~\ref{tab:tnx_success_model_input} lists the input parameters to our success rate model. Specifically, $RK$, $P_{RK}$ should be same-sized lists specifying the keys that read requests may contain, and the probability of such keys being requested. Similar for $WK$ and $P_{WK}$ to describe write requests.

\begin{table}[H]
\caption{Input to transaction success rate model.}
\label{tab:tnx_success_model_input}
\begin{tabular}{l|l|l}
Variable & Description & Unit/Value \\ \hline
$RP$ & Probability of read request in single transaction & Float $\in [0, 1]$ \\
$WP$ & Probability of write request in single transaction & 1 - RP \\
$RK$ & List of possible keys that read request may contain & List of integers \\
$P_{RK}$ & Probability of keys in read requests & List of floats sum to 1 \\
$WK$ & List of possible keys that write request may contain & List of integers \\
$P_{WK}$ & Probability of keys in write requests & List of floats sum to 1 \\
$BS$ & Number of transactions in a block & count
\end{tabular}
\end{table}

\noindent\textbf{Model.} Consider two same-block transactions $(a, b)$, with $a$ ordered before $b$. We start by defining the failure probability of $b$:

\textit{Scenario 1:} Read-Write Failure. The probability for read-write is $RP*WP$ and probability for read-write key conflict is $\sum_{i\in WK} P_{RK}(i)P_{WK}(i)$. Overall, the probability of a Read-Write Failure is

\begin{equation}\label{eq:prwf}
P_{RWFail} = RP*WP\sum_{i\in WK} P_{RK}(i)P_{WK}(i)
\end{equation}

\textit{Scenario 2:} Write-Read Failure. The probability for write-read is $WP*RP$ and probability for read-write key conflict is $\sum_{i\in RK} P_{WK}(i)P_{RK}(i)$. Overall, the probability of a Write-Read Failure is

\begin{equation}\label{eq:pwrf}
P_{WRFail} = WP*RP\sum_{i\in RK} P_{WK}(i)P_{RK}(i)
\end{equation}

\textit{Scenario 3:} Write-Write Failure. The probability for both-write is $WP^2$ and probability for two write key conflict is $\sum_{i\in WK} P_{WK}(i)^2$. Overall, the probability of a write-write failure is

\begin{equation}\label{eq:pwwf}
P_{WWFail} = WP^2\sum_{i\in WK} P_{WK}(i)^2
\end{equation}

Note that in scenario 1 the summation is over $i\in WK$, and in scenario 2 the summation is over $i \in RK$. This is to differentiate different key ranges that read and write requests may take.

Combining the three aforementioned scenarios, for any ordered pair of transactions $(a, b)$, the probability of $b$ to fail is

\begin{equation}\label{eq:psf}
P_{bFail} = P_{RWFail} + P_{WRFail} + P_{WWFail}
\end{equation}

Note that equation~\ref{eq:psf} is computed based on the input read/write and key distributions. Next, we consider the number of transactions within the block to derive the block-level success rate of transactions.

Suppose the transactions in the block is numbered as $1,2,\dots,BS$. Then, the $k^{th}$ transaction's probability of success can be computed as the probability of non-failure with all of its $k-1$ preceding transactions:

\begin{equation}\label{eq:kts}
P_{kTxnSuccess}(k) = (1 - P_{bFail})^{k-1} = (1 - P_{RWFail} - P_{WRFail} - P_{WWFail})^{k-1}
\end{equation}

Note that equation~\ref{eq:kts} strongly relies on the assumption of i.i.d. transactions.

Treating each transaction's success as an indicator random variable with positive probability given by equation~\ref{eq:kts}, the expected number of successful transactions in a block is given by:

\begin{equation}\label{eq:expected_success_tnx}
E[\text{\# successful transactions in block}] = \sum_{k=1}^{BS} P_{kTxnSuccess}(k) = \sum_{k=1}^{BS}(1 - P_{RWFail} - P_{WRFail} - P_{WWFail})^{k-1}
\end{equation}

\subsection{Empirical study}

To validate our theoretical model, we implemented a framework to run simulated experiments: peers with predefined access patterns (i.e. $RP$, $WP$, $RK$, $P_{RK}$, $WK$, $P_{WK}$) submit transactions to a simulated block server with fixed block size (i.e. $BS$), then we count the number of successful transactions within each simulated block; finally we compute the overall transaction success rate. We chose to implement the simulation framework instead of leveraging existing platforms (e.g. HyperLedger Fabric) because our simulation framework has more flexible usages.

\noindent\textbf{Background: Zipfian distributions.} Most of our experiments assume a ranged Zipfian probability distribution (e.g. $P_{WK}$ and $P_{RK}$). In our used case, a ranged Zipfian distribution is characterized by three parameters: $range$, $\alpha$, and $reverse$.

In a non-$reversed$ distribution, the probability of an integer $n \in [1, range]$ is given by:

\begin{equation}\label{eq:zipf}
P(n) = \frac{\alpha^{-n}}{\sum_{i=1}^{range} \alpha^{-i}}
\end{equation}

In $reversed$ version, the probability is given by:

\begin{equation}\label{eq:zipf-reversed}
P(n) = \frac{\alpha^{-(range + 1 - n)}}{\sum_{i=1}^{range} \alpha^{-i}}
\end{equation}

\begin{figure}[!htp]
    \centering
    \includegraphics[scale=0.55]{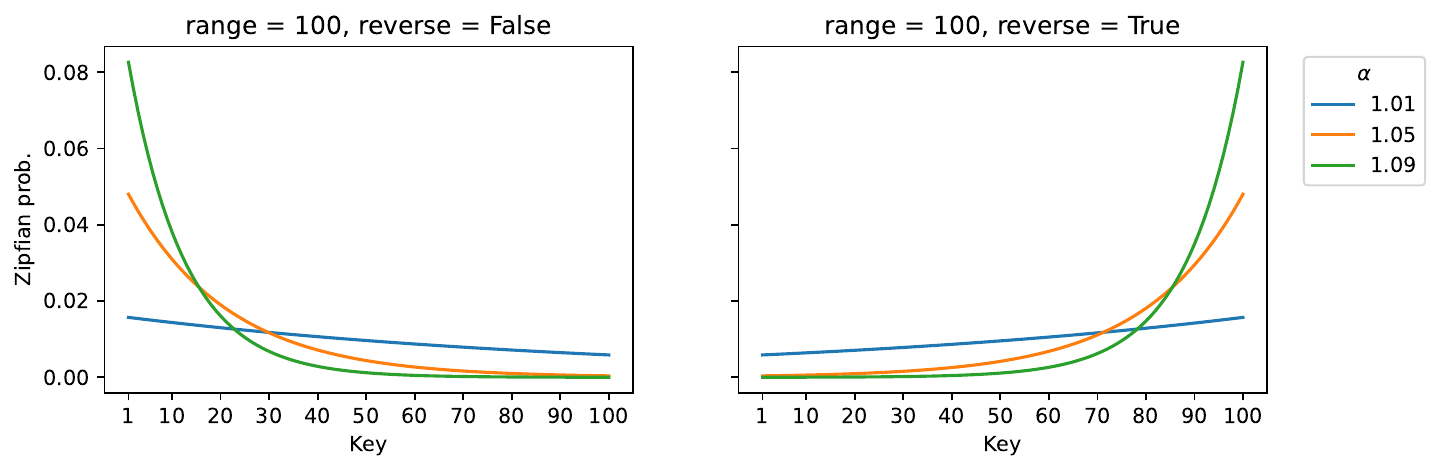}
    \caption{Ranged Zipfian distribution with $range=100$, non-reversed (left) and reversed (right), and various $\alpha$ values showing as multiple lines in each plot.}
    \label{fig:zipf}
\end{figure}

Figure~\ref{fig:zipf} shows the probability distribution function of ranged Zipfian distributions with various parameters. Specifically, the $\alpha$ parameter controls the skewness of the distribution: larger $\alpha$ increases the likelihood of generating same values.

\subsubsection{Case study 1: simple only-write clients with ranged Zipfian keys.}

In this experiment, we consider all clients only submit write-typed transactions. Thus, $RP=0$ and $WP=1$. The transaction-requested keys follow ranged Zipfian distributions. We varied the $\alpha$, $range$, and $BS$ values of our simulator and collected measured success rate for 1000 operations. For every combination of $\alpha$, $range$, and $BS$ values, we ran 50 simulations and keep the 1, 50, 99 percentile success rates to show the tail and average measurements. We then compare the measured success rates with the theoretical model.

\begin{figure}[!htp]
    \centering
    \includegraphics[scale=0.45]{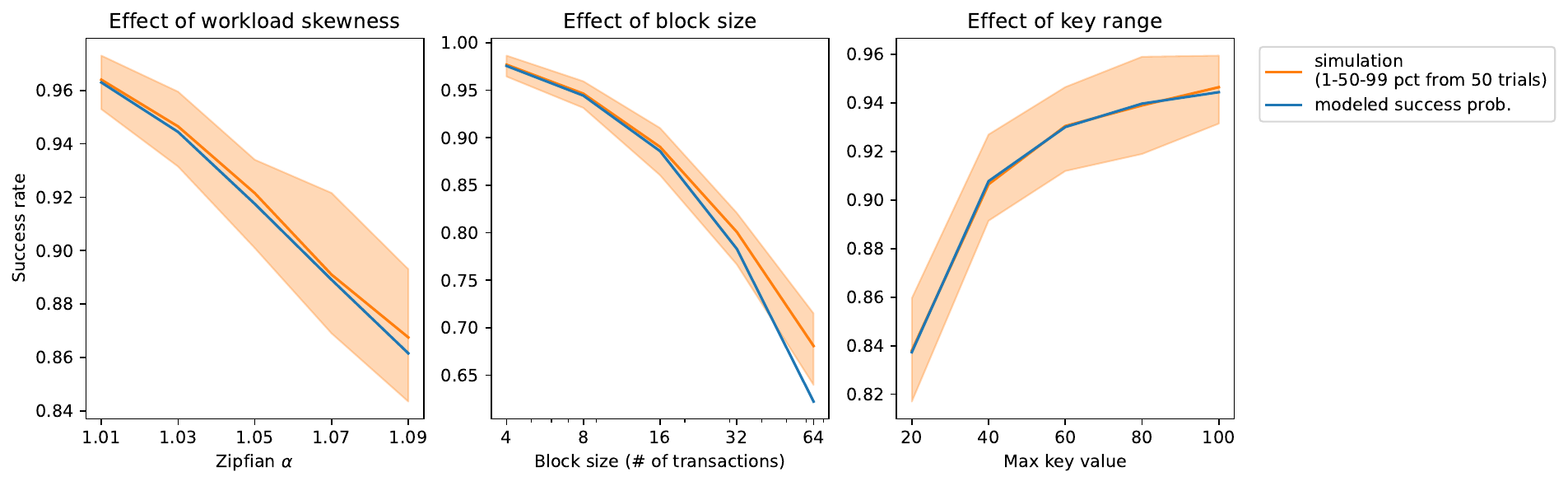}
    \caption{Simulation and model-derived success rate for all-write clients, in varying $\alpha$ (left), $BS$ (middle), and $range$ (right) settings. The shaded orange region indicates the 1-99 percentile from 50 simulation trials, and the orange line shows the median simulation result. $\alpha$ is fixed to 1.03 in middle \& right, $BS$ is fixed to 8 in left \& right, and $range$ is fixed to 100 in left \& middle.}
    \label{fig:all-write}
\end{figure}

Figure~\ref{fig:all-write} shows the comparison. The detailed figure setting can be found in the caption. Overall, the plots show that (a) our theoretically modeled success rate is fairly close to simulation measurements, and (b) both theoretically modeled success rate and simulation measurements agree with intuition: the left figure shows that increased skewness in key distribution causes success rate to drop; the middle figure shows increased block size causes key conflict likelihood to rise, thus overall success rate to drop; the right figure shows increased key range causes the key distribution to be less skewed (as in equation~\ref{eq:zipf}), thus improves the overall success rate.

\subsubsection{Case study 2: read-write clients.} In this experiment, we consider clients operates under a read-write access pattern: first, the client submits a read request to a key following ranged Zipfian distribution; then, the client submits a write request to the same key. After a pair of (read, write) requests are processed, the client selects a new key and start over. If a transaction failed, the client will resubmit the same transaction until succeeds, then submit new ones.

To derive the theoretical modeled success rate, we consider $RP=WP=0.5$, and $RK=WK$, $P_{RK}=P_{WK}$. Similar to case study 1, we run repeated simulations under different ($\alpha$, $BS$, $range$) combinations and record 1, 50, and 99 percentiles.

\begin{figure}[!htp]
    \centering
    \includegraphics[scale=0.45]{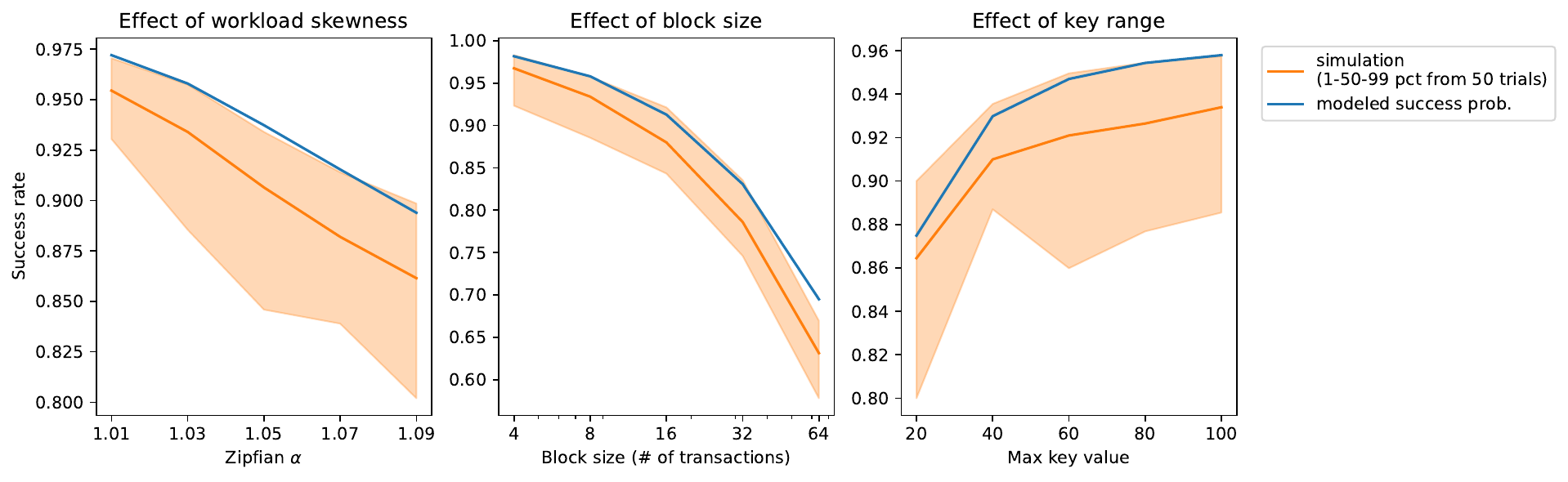}
    \caption{Simulation and model-derived success rate for read-write clients, in varying $\alpha$ (left), $BS$ (middle), and $range$ (right) settings. The shaded orange region indicates the 1-99 percentile from 50 simulation trials, and the orange line shows the median simulation result. $\alpha$ is fixed to 1.03 in middle \& right, $BS$ is fixed to 8 in left \& right, and $range$ is fixed to 100 in left \& middle.}
    \label{fig:read-write}
\end{figure}

Figure~\ref{fig:read-write} shows the simulated and theoretically modeled success rates for read-write clients, with varying $\alpha$, $BS$, and $range$ settings (left, middle, and right subplots). The details can be found in the figure caption.

Similar to the all-write case study, figure~\ref{fig:read-write} agrees with our intuition. However, in this case, our theoretically modeled success rate is closer to the best-case simulation measurement (i.e. 99-percentile) than to the median and tail. This is because our theoretical model assumes $RP=WP=0.5$, while in the simulations the read/write ratio is different due to clients retrying failed transactions. In general, write requests are more likely to fail since 2 of the 3 failing scenarios (i.e. read-write, write-read, write-write) will cause a write request re-submission. Nevertheless, our theoretical modeled success rate is within the observation range and around at most $3\%$ away from median case.

\subsubsection{Case study 3: clients with different read/write key access pattern.}\label{sec:rw-diff-analysis} 
\begin{figure}[!htp]
    \centering
    \includegraphics[scale=0.55]{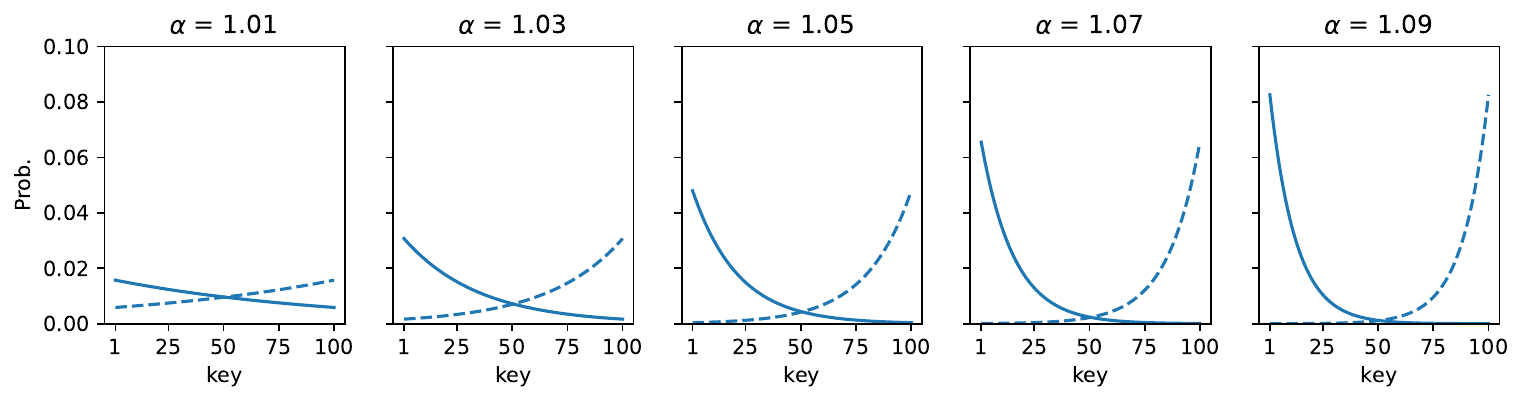}
    \caption{Non-reversed (solid) and reversed (dashed) ranged zipfian distribution for different $\alpha$ values.}
    \label{fig:zipf-reverse}
\end{figure}

In this case study, we consider clients with different read/write key distributions, specifically, the read and write key distributions are ranged Zipfian distributions with identical $\alpha$ and $range$, but the write key distribution is reversed (i.e. equation~\ref{eq:zipf-reversed}). Figure~\ref{fig:zipf-reverse} shows the read and write key distribution for different $\alpha$ values. Read key distributions are shown as solid line and write ones are dashed.

In this experiment, we focus on varying $RP$ and $\alpha$. We skip the plots for varying other parameters since they gave similar trend as dicussed in previous case studies.

\begin{figure}[!htp]
    \centering
    \includegraphics[scale=0.45]{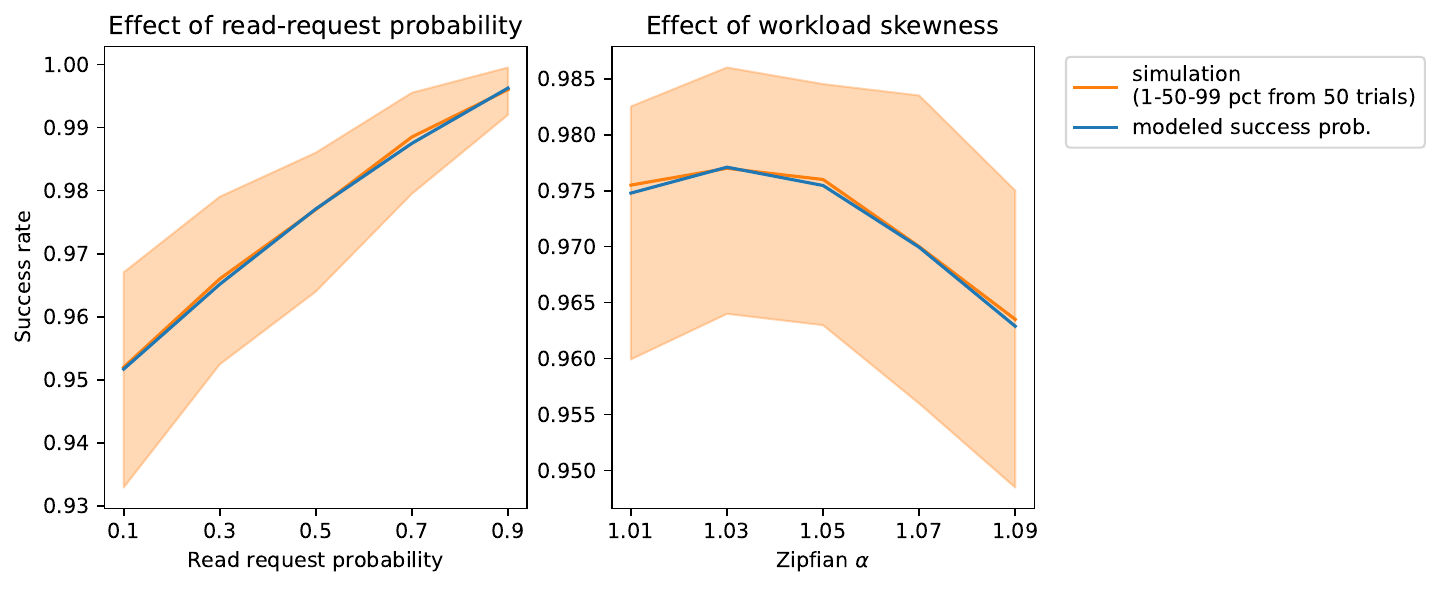}
    \caption{Simulation and model-derived success rate for clients with different read and write access pattern, in varying $RP$ (left) and $\alpha$ (right) settings. The shaded orange region indicates the 1-99 percentile from 50 simulation trials, and the orange line shows the median simulation result. $\alpha$ is fixed to 1.03 in left, $RP$ is fixed to 0.5 in right. $BS$ is fixed to 8, and $range$ is fixed to 100 in both left and right figures.}
    \label{fig:rw-diff}
\end{figure}

Figure~\ref{fig:rw-diff} shows the simulated and theoretically modeled success rates for clients with reversed read and write key distributions, with varying read request probability (left) and Zipfian skewness parameter (right). We are able to observe new trends:
\begin{itemize}
    \item In figure~\ref{fig:rw-diff} left, we see that success rate increases as read request probability increases. This agrees with our intuition such that read-read key conflicts will not cause transaction failures.
    \item In figure~\ref{fig:rw-diff} right, we see that key distribution skewness does not have a linear impact on success rate; rather, as $\alpha$ increases, both simulated and modeled success rates first increase, then decrease. Our hypothesis for such observation is: as $\alpha$ increases, the likelihood for write-write failure increases, but the skewed region (i.e. "hot keys") for read and write will be further separated (closer to $1$ and $100$ respectively, as shown in figure~\ref{fig:zipf-reverse}), causing less read-write and write-read conflicts.

\begin{table}[H]
\caption{Write-write conflict probability and read-write curve overlap area for different $\alpha$ values in clients with reversed read and write probability distributions.}
\label{tab:rw-diff-stats}
\begin{tabular}{l|lllll}
$\alpha$ & 1.01 & 1.03 & 1.05 & 1.07 & 1.09 \\ \hline
$P_{WWFail}$ & 0.0108 & 0.0164 & 0.0248 & 0.0339 & 0.0431 \\
R/W overlap area & 0.75 & 0.36 & 0.16 & 0.07 & 0.03
\end{tabular}
\end{table}

    To validate our explanation hypothesis, for multiple $\alpha$ values, we compute (1) the write-write key conflict probability using equation~\ref{eq:pwwf}, and (2) the overlapping area of read-write probability distribution curves as in figure~\ref{fig:zipf-reverse} using the composite trapezoidal rule. The results are shown in table~\ref{tab:rw-diff-stats}. As $\alpha$ increases, $P_{WWFail}$ increases, and R/W overlap area decreases. Such observation proves our hypothesis.
\end{itemize}

Note that the theoretical modeled success rates are close to the median simulation results in all parameter configurations, showing the model's accuracy.

\subsubsection{Conclusion on transaction success rate modeling.}
To summarize our case study efforts, we show the accuracy of our theoretical transaction success rate model by comparing to simulation-collected measurement results. In addition, our theoretical model can successfully demonstrate the non-trivial effect of complex parameter settings on success rate (e.g. section~\ref{sec:rw-diff-analysis}).

Besides simulations, we also developed methodologies to check for failed transactions in HyperLedger Fabric, described in Appendix~\ref{sec:hpf-tnx-failure}.

\section{Conclusion}

In this project, we focus on the modeling of block creation, average transaction latency and transaction success rate. And we evaluate our average transaction latency model on open-source blockchain systems hyperleger fabric and measure the resource usage (e.g. CPU utilization and IO). Finally, we validate our transaction success rate model using simulation experiments. 
\bibliographystyle{ACM-Reference-Format}
\bibliography{References}

\appendix
\section{source code analysis of block creation in hyperledger fabric}

Here is the function call tree when orderer persists newly created block

\begin{enumerate}
    \item In {\footnotesize \texttt{func (c *Chain) apply(ents []raftpb.Entry)}} \href{https://github.com/hyperledger/fabric/blob/v2.2.10/orderer/consensus/etcdraft/chain.go#L1041}{[link]}, the orderer node invokes\\ {\footnotesize \texttt{c.writeBlock(block, ents[i].Index)}}
    \item In {\footnotesize \texttt{func (c *Chain) writeBlock(block *common.Block, index uint64) }}\href{https://github.com/hyperledger/fabric/blob/v2.2.10/orderer/consensus/etcdraft/chain.go#L849}{[link]}, the orderer node invokes\\ {\footnotesize\texttt{c.support.WriteBlock(block, m)}}
    \item In {\footnotesize\texttt{func (bw *BlockWriter) WriteBlock(block *cb.Block, encodedMetadataValue []byte)}}\href{https://github.com/hyperledger/fabric/blob/v2.2.10/orderer/common/multichannel/blockwriter.go#L174}{[link]}, the orderer node invokes {\footnotesize\texttt{bw.commitBlock(encodedMetadataValue)}}. Notice that the {\footnotesize \texttt{bw.committingBlock}} ensures blocks are committed in a serial order. 
    \item In {\footnotesize \texttt{func (bw *BlockWriter) commitBlock(encodedMetadataValue []byte)}} \href{https://github.com/hyperledger/fabric/blob/v2.2.10/orderer/common/multichannel/blockwriter.go#L180}{[link]}, the orderer invokes\\ {\footnotesize \texttt{err := bw.support.Append(bw.lastBlock)}}
    \item In {\footnotesize \texttt{func (fl *FileLedger) Append(block *cb.Block) error}} [\href{https://github.com/hyperledger/fabric/blob/v2.2.10/common/ledger/blockledger/fileledger/impl.go#L109}{link}], the orderer invokes\\ {\footnotesize \texttt{err := fl.blockStore.AddBlock(block)}}
    \item In {\footnotesize \texttt{func (store *BlockStore) AddBlock(block *common.Block) error}} [\href{https://github.com/hyperledger/fabric/blob/v2.2.10/common/ledger/blkstorage/blockstore.go#L47}{link}], the orderer invokes\\ {\footnotesize \texttt{	result := store.fileMgr.addBlock(block)}}. The {\small \texttt{elapsedBlockCommit}} measures \textit{block commit time}. 
    \item In {\footnotesize \texttt{func (mgr *blockfileMgr) addBlock(block *common.Block) error}}, most of the commit time is spent in appending file \href{https://github.com/hyperledger/fabric/blob/v2.2.10/common/ledger/blkstorage/blockfile_mgr.go#L318}{[link]} and save the index \href{https://github.com/hyperledger/fabric/blob/v2.2.10/common/ledger/blkstorage/blockfile_mgr.go#L352}{[link]}in the database. 
\end{enumerate}

Here is the logic of orderer sends the newly persisted block
\begin{enumerate}
    \item In {\footnotesize \texttt{func (mgr *blockfileMgr) addBlock(block *common.Block) error}}, the orderer invokes \\{\footnotesize \texttt{	mgr.updateBlockfilesInfo(newBlkfilesInfo)}}\href{https://github.com/hyperledger/fabric/blob/v2.2.10/common/ledger/blkstorage/blockfile_mgr.go#L359}{[link]} after persists the new block. 
    \item In {\footnotesize \texttt{func (mgr *blockfileMgr) updateBlockfilesInfo(blkfilesInfo *blockfilesInfo)}}\href{https://github.com/hyperledger/fabric/blob/v2.2.10/common/ledger/blkstorage/blockfile_mgr.go#L487}{[link]}, the orderer invokes\\ {\footnotesize \texttt{	mgr.blkfilesInfoCond.Broadcast()}} to wake goroutine that waits to send new blocks. 
    \item In {\footnotesize \texttt{func (itr *blocksItr) waitForBlock(blockNum uint64) uint64}} \href{https://github.com/hyperledger/fabric/blob/v2.2.10/common/ledger/blkstorage/blocks_itr.go#L37}{[link]}, this goroutine waits at \\{\footnotesize \texttt{		itr.mgr.blkfilesInfoCond.Wait()}} is waken. 
    \item In {\footnotesize \texttt{func (itr *blocksItr) Next() (ledger.QueryResult, error)}}\href{https://github.com/hyperledger/fabric/blob/v2.2.10/common/ledger/blkstorage/blocks_itr.go#L82-L83}{[link]}, this iterator returns the new deserialized block.
    \item In {\footnotesize \texttt{func (h *Handler) deliverBlocks(ctx context.Context, srv *Server, envelope *cb.Envelope)}} \href{https://github.com/hyperledger/fabric/blob/v2.2.10/common/deliver/deliver.go#L299}{[link]}, the other goroutine of orderer gets this new block and sends this new block  by invoking {\footnotesize \texttt{err := srv.SendBlockResponse}} \href{https://github.com/hyperledger/fabric/blob/v2.2.10/common/deliver/deliver.go#L332}{[link]}.
\end{enumerate}

Observation: \textit{committing blocks and sending blocks are two goroutines that run in parallel. }

The logic for peer to receive blocks:  \href{https://github.com/hyperledger/fabric/blob/v2.2.10/internal/pkg/peer/blocksprovider/blocksprovider.go#L116}{[link]}

The logic for peer to commit blocks:  \href{https://github.com/hyperledger/fabric/blob/v2.2.10/core/ledger/kvledger/kv_ledger.go#L445}{[link]} and there are four parts. 
\begin{itemize}
    \item validate state
    \item commit pvtdata and block to storage
    \item commit txns in block to state database
    \item commit txns to history database
\end{itemize}

\section{transaction failures in HyperLedger Fabric}
\label{sec:hpf-tnx-failure}

In HyperLedger Fabric, it is possible to check for failed transactions entirely on block server side, without relying on peers to validate. This allows transaction validation in both \texttt{open-loop} and \texttt{closed-loop} modes, since \texttt{open-loop} does not wait for peer validation response.

We check for failed transactions by parsing each received block in block server to retrieve each transaction's request type (read or write) and request key:

\begin{enumerate}
    \item Existing logic in block server: in \texttt{block-server.js}, the \texttt{getChannel(channelName).then((network)...} function contains code for actively listening for newly created blocks. When the listener discovers an event, the block is obtained by \texttt{const block = event.blockData}.
    \item Then, transactions are saved as list in \texttt{block.data.data}. For example, we can extract the first transaction by \texttt{const tnx = block.data.data[0]}.
    \item We next extract the actions within the transaction by \texttt{tnx.payload.data.actions}. Usually each transaction only contains one action, thus we obtain the action by \texttt{const action = tnx.payload.data.actions[0]}.
    \item We extract the input by \texttt{const input = action.payload.chaincode\_proposal\_payload.input}.
    \item Finally, from the input, we extract the arguments by \texttt{const args = input.chaincode\_spec.input.args}.
    \item \texttt{args} will be a three-element list of bytes, representing the action type, key, and value respectively. Note that the bytes will need to be converted to string.
\end{enumerate}

\begin{figure}[!htp]
    \centering
    \includegraphics[scale=0.45]{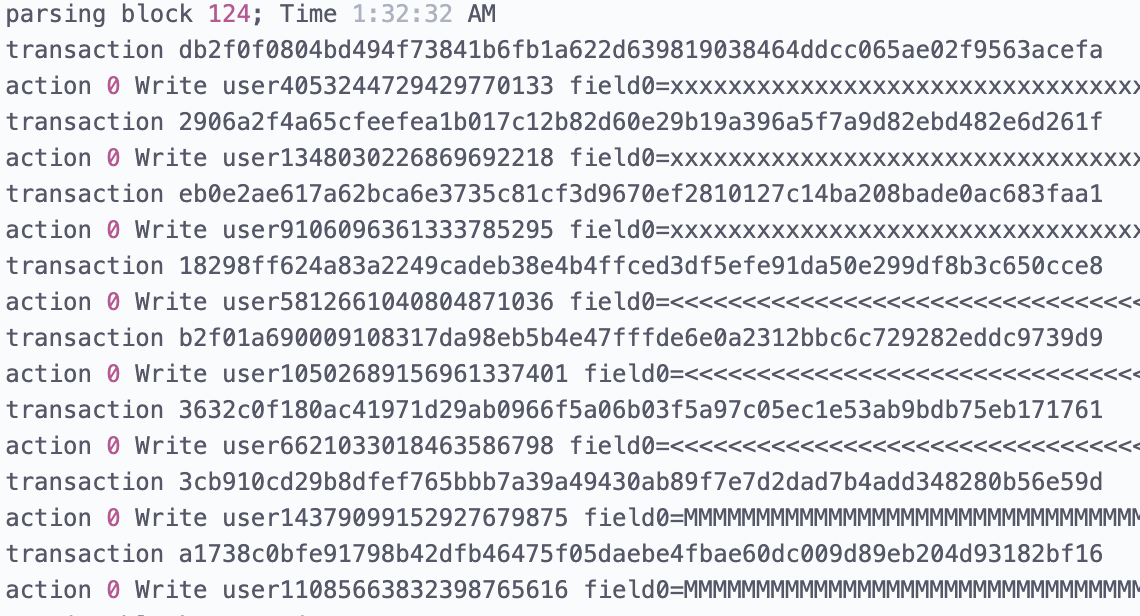}
    \caption{A sample block in the kv-store application provided by HyperLedger Fabric.}
    \label{fig:hlf-tnx}
\end{figure}

Figure~\ref{fig:hlf-tnx} shows a sample block's transaction types and keys parsed using the aforementioned technique. Note that in HyperLedger Fabric's kvstore application, the keys are the concatenation of string \texttt{"user"} and the SHA256 hash of a Zipfian (or other distributions per user request) generated number. The write values are the strings begin with \texttt{"field0..."} which are not relevant to failure analysis.

\end{document}